# Hydrogen permeability prediction in palladium alloys and virtual screening of B2-phase stabilized $Pd_{(100-x-y)}Cu_xM_y$ ternary alloys using machine learning

Eric Kolor[1,*], Edoardo Magnone[2], Muhammad Harussani Moklis[1], Md. Rubel[1], Sasipa Boonyubol[1], Koichi Mikami[1], Jeffrey S. Cross[1]

[1] Energy Science and Engineering, Department of Transdisciplinary Science and Engineering, Institute of Science Tokyo, 2–12–1, Ookayama, Meguro-ku, Tokyo 152–8550, Japan

[2] Department of Chemical and Biochemical Engineering, Dongguk University, 30, Pildong-ro 1 gil, Jung-gu, Seoul, 04620, South Korea

* Corresponding author (kolor.k.aa@m.titech.ac.jp)

## Abstract

We present a forward prediction material screening framework designed to discover Pd–Cu alloys with improved B2 phase stability, thereby unlocking simultaneous hydrogen generation and utilization. First, we trained CatBoost regressors with literature-derived Pd-alloy data to predict hydrogen permeability from composition and testing conditions. We evaluated fractional, composition-based, and physics-informed descriptors, individually and in combination, and showed that sequential Pearson filtering and fold-wise Shapley Additive Explanations (SHAP)-based recursive feature elimination with cross-fold aggregation reduced errors while controlling complexity. Guided by the one-standard-error rule, a narrower domain-informed set of 13 features provided the best accuracy parsimony trade-off ($R^2 = 0.81$), only 0.01 below the maximum $R^2$ achievable with triple the number of features. SHAP analysis indicated that high permeability is promoted by elevated temperature, lattice expansion relative to Pd, atomic size mismatch, and favorable mixing tendencies. Second, the selected model was applied to screen $Pd_{(100-x-y)}Cu_xM_y$ spanning 16 co-dopants for B2 stabilization. For each co-dopant system, we obtained the Pareto set of compositions that minimize Pd content and Miedema heat of formation and maximize the permeability, then picked three compounds, including that with the highest predicted permeability, the lowest Miedema heat of formation, and the lowest Pd content. With a final filter considering co-dopant concentration for single-phase Pd–M solution formation, we recommend $Pd_{48.48}Cu_{43.00}Y_{8.52}$, $Pd_{49.08}Cu_{42.45}Sc_{8.47}$, $Pd_{56.09}Cu_{33.70}La_{10.21}$, and $Pd_{52.68}Cu_{40.44}Mg_{6.88}$ for experimental validation. We predict those alloys to exhibit permeabilities +1.7 – +1.9 times higher than B2–$Pd_{60}Cu_{40}$ (all in wt.%). Our framework provides plausible experimental targets and a scalable pathway for designing stable, high-temperature, $H_2$-selective Pd-alloy membranes.

## 1. Introduction

Over 150 years after Thomas Graham's report of hydrogen occlusion in palladium (1866), truly cost-competitive metallic membranes that withstand harsh industrial service remain an open challenge [1]. The decarbonization potential of green hydrogen ($H_2$) increases the urgency for durable, selective, and economical $H_2$-separation technologies. Palladium (Pd) alloys possess uniquely active $H_2$-dissociating surfaces and a lattice that supports rapid, thermally activated proton transport, giving them advantages over group IV/V and Ni-based alloys. Membrane processes are also modular and require lower initial capital expenditures than pressure-swing adsorption, offering a path to process intensification, provided key bottlenecks are addressed.

A central obstacle is reducing Pd content in order to decrease the membrane cost without sacrificing performance. Palladium–copper (Pd–Cu) membranes are a pragmatic compromise: Cu lowers cost, improves corrosion resistance, thermal expansion matching, tensile strength, sulfur-poisoning and embrittlement resistance [1]. However, near-equiatomic Pd–Cu alloys with an ordered body-centered cubic (*bcc*) structure, or B2, the composition associated with the highest theoretical permeability and catalytic activity, undergoes a B2 → B2 + disordered face-centered cubic (*fcc*) or A1 → A1 transition upon $H_2$ uptake and heating, with substantial A1 formation by ~ 400–598 °C [2–5]. Maintaining stable B2-like surface chemistry and high permeability upon temperature cycling would benefit integrated applications such as catalytic membrane reactors and reformed methanol fuel cells [6–8].

We hypothesize that B2 stabilization can be extended into ternaries by judicious addition of a third metal, M, chosen to balance activity, stability, and mechanical properties. A third element stabilizes B2–Pd–Cu by lowering the free energy of the ordered state relative to disorder (B2 → A1), through a combination of stronger site-specific bonding, size-strain accommodation, reduced configurational entropy gain upon disordering, and non-trivial site-substitution effects, rather than simply by making the ternary formation enthalpy more negative [9]. This suggests a large design space: which elements promote B2 stability, prefer Pd or Cu substitution, how much should be added, and what trade-offs emerge (sulfur tolerance, hydrogen dissociation, ductility, CO resistance, etc.)? Prior works illustrated both opportunities and gaps. High-throughput first-principles calculations have been used to evaluate substitutionally random Pd-rich *fcc* $Pd_{100-x-4}Cu_xM_4$ (M ∈ {Ag, Au, Nb, Ni, Pt, Rh, Ru, Ta, Ti, V, Zr}) alloys for high surface chemical activity, surface segregation, and sulfur poisoning resistance [10,11]. Patent by Benn et al. suggested using Hume-Rothery rules and density functional theory (DFT) based heat of formation criteria that the B2-phase in $Pd_{[35-55]}Cu_xM_{[1-15]}$ alloys could be stabilized using a *bcc* metal M ∈ {Fe, Cr, Nb, Ta, V, Mo, W} [12]. Patents and papers by U.S. NETL researchers used DFT-calculated heat of formation as a criterion to screen 37 metallic elements which could stabilize $B2-Pd_{8-x}Cu_8M_x$ and suggested M ∈ {Al, Ga, Hf, La, Mg, Sc, Ti, Y, Zn, Zr} as strong candidates [13–16]. They additionally reported experimentally that Mg might be the best B2 phase stabilizer at 400 ºC [16]. Experimental membrane-based Japanese patents have also reported dilute additions of $M \in \{Al, Ga, In\}$ to form $Pd_{[41-50]}Cu_{[48-58.8]}M_{[0.2-2.0]}$ that maintain permeability > $1.0 \times 10^{-8}$ $mol \cdot m^{-1} \cdot s^{-1} \cdot Pa^{-0.5}$ above 600 °C [7,8]. More recently, Horikawa et al. have succeeded in improving by 1.4 times (compared to the peak permeability of B2–$Pd_{60}Cu_{40}$ (wt%)) the permeability coefficient of B2–$Pd_{60}Cu_{40}$ (wt%) at 300°C by increasing Pd content by 1% and applying a controlled annealing treatment [2]. They further demonstrated using CALculation of PHase Diagram (CALPHAD), DFT and experimental testing that dilute Cu sites substitution in $Pd_{61}Cu_{39}$ (wt%) by Mn and Al is a promising avenue for a more stable B2 phase at T ≥ 450 °C [17]. Yet, outside a few case studies [7,8,17], the expected permeability ranges of B2-stabilized Pd–Cu–M alloys remain largely unmapped because extensive syntheses, characterizations, testing, and *ab initio* simulations are required.

Material informatics offers a complementary route. Recent successes in metallic and Metal-Organic-Framework (MOF)-based membranes [18,19], alloys [20,21], and catalysis [22,23], show that data-driven models can accelerate property prediction and guide targeted experiments. Yang et al. recently proposed a combined DFT and machine learning workflow to evaluate Pd-alloy/gas interactions [19]. Using 30 features spanning simple descriptors (e.g., atomic radius) and DFT-derived quantities (e.g., adsorption energies and binding energies), they reported eXtreme Gradient Boosting (XGBoost) as the strongest selectivity predictor ($R^2$ = 0.96) and Gradient Boosting Decision Trees (GBDT) as the strongest permeance predictor ($R^2$ = 0.99) [19], with the high scores likely reflecting the homogeneity of the DFT-optimized structures.

In this work, we addressed the absence of reliable data-driven tools for predicting hydrogen permeability in Pd alloys and for screening large pools of Pd alloy candidates. First, we constructed a standardized dataset of experimentally observed non-composite Pd alloy membranes. Second, we trained an interpretable small-data-friendly algorithm viz. Categorical Boosting (CatBoost) regressor to model the non-linear composition–testing condition–permeability relationship. Lastly, we used the model to map composition-permeability domains for 16 B2-stabilizing elements, revealing Pd-lean B2-phase-stabilized Pd–Cu–M candidates suitable for experimental validation.

## 2. Methodology

### 2.1. Material dataset preparation

We manually compiled a tabular dataset enriched with metadata, covering experimentally observed dense crystalline Pd-alloy membranes and their operating conditions. To construct the dataset, we retained 71 sources that provided sufficiently complete experimental information, ensuring minimal missing values aside from the lattice parameter. These sources included peer-reviewed journal articles, technical reports, and patents, from which we compiled 333 distinct alloy compositions. We converted each alloy formula to atomic fractions and, whenever available, prioritized the experimentally determined compositions from solid-state characterization over the nominal targeted ratios. The analysis focused exclusively on pure $H_2$ permeation experiments in non-composite membranes, namely planar and tubular configurations without any porous support underlying the active diffusion area.

For each membrane sample, we extracted the experimental thickness, diffusion temperature, pressure differentials ( $\Delta P^n = P^n_{feed} - P^n_{permeate}$) at which hydrogen fluxes ($J_{H_2}$) were measured, and the pressure exponent ($n$), which characterizes the rate-limiting step in hydrogen diffusion. Most flux data were obtained by digitizing $J_{H_2} - \Delta P^n$ plots using WebPlotDigitizer [24]. Lattice constants were either (*i*) taken directly when reported, (*ii*) calculated using non-linear least-squares refinement with UnitCell software [25,26], or (*iii*) estimated using Vegard's law without bowing correction when explicit or sufficient XRD data were unavailable. For binary Pd–Cu alloys, cubic lattice constants were instead estimated using the Vegard's linear regression equations of Al-Mufachi et al. [27], applied to the relevant region of the Pd–Cu phase diagram [3,5,28].

Hydrogen permeability of dense metallic membranes is conventionally quantified by measuring hydrogen fluxes ( $J_{H_2}$ ) under varying pressure differentials ($\Delta P^n$) at fixed temperature, followed by linear regression of $J_{H_2}$ versus $\Delta P^n$ . The gradient of this regression line yields the permeance (permeability normalized by thickness) according to Richardson's equation [1,29], which is often reported as the permeability coefficient. To maximize dataset size and capture the full range of experimental flux values, we instead computed pointwise permeability directly from reported fluxes and pressure differentials according to Eq. (1):

$$\text{Permeability } (\phi) = \frac{J_{H_2} \times \text{Thickness}}{\Delta P^n} \quad (1)$$

The pressure exponent ($n$) was adopted from the values reported by the original authors, typically the exponent that maximized the coefficient of determination ($R^2$) in the $J_{H_2} - \Delta P^n$ regression. Although pointwise permeability is more sensitive to experimental noise than slope-derived values, it allowed us to retain all available datapoints, thereby increasing the number of observations per membrane and enhancing the statistical robustness of the machine learning models.

## 2.2. Data processing

We performed several outlier detection analyses before removing records with $n > 0.5$ . Although these values are physically meaningful, they appeared as rare outliers in the corpus and did not provide sufficient frequency to support reliable model generalization. Consequently, the scope of our predictions strictly applies to $n = 0.5$ , *i.e.*, conditions that obey Sieverts' law. In this regime, hydrogen solubility in the bulk alloy is proportional to the square root of its partial pressure, implying that the rate-limiting step of permeation is diffusion through the bulk metal. Consequently, $n$ becomes a zero-variance variable, thus carried no information and was not used as a feature. Although the distribution of the permeability is left-skewed (Fig. S1, Supplementary Material) due to outlier data points that we deemed scientifically plausible, we confirmed that removing those data points is deleterious for further prediction. For example, we conserved copper-rich binary membrane with zero permeability [28], and iron-rich alloys with ~$10^{-14}$ mol·m$^{-1}$·s$^{-1}$·Pa$^{-0.5}$ permeability [30]. We confirmed that general methods to handle skewed distribution (Yeo-Johnson and Box-Cox power transformations) have failed even after removing low-magnitude permeability points. In this work, we predicted $\log(1 + Permeability)$ where $Permeability$ is the pointwise permeability, but we always report the results on the original scale. Features were systematically robust-scaled to temper outliers and mitigate scale variability, although shallow gradient boosted decision trees models such as CatBoost do not necessarily require any feature scaling nor target values transformations (see '3. Data scaling' section in the Supplementary Material for details).

### 2.3. Composition-aware data partitioning with column-similarity stratification

We grouped identical alloy formulas and stratified the obtained groups by column similarity in the periodic table. Grouping places all observations of the same canonical composition in one partition, enforcing mutual exclusivity and preventing leakage across training, cross-validation, and held-out test sets.

Stratification exploits a column-similarity encoding using ascending IUPAC group numbers (e.g., $Pd_{91.12}Ti_{8.88}$ → G4–G10) to reflect shared valence-electron chemistry, preserve composition-family frequencies in train/test splits, and reduce covariate shift in a regression setting. Rare strata with fewer than 20 instances were merged into the "Others" bin (Fig. 1). Pure palladium (G10) served as a calibration reference during training and was excluded from test scoring. Splits of 80/20, 85/15, and ca. 95/5, in addition to distribution conservation across the various data sizes, are summarized in Table 1 and Fig. 1, respectively.

### 2.4. Feature engineering

We evaluated four descriptor families and their combinations. "Exp" captures experimental conditions (see Sec. 2.1) and a physics-motivated lattice-misfit term given by:

$$a_{ss} \,/\, a_{Pd} = (a_{alloy} - a_{Pd}) \,/\, a_{Pd} \quad (2)$$

which quantifies Pd lattice expansion or contraction upon alloying. "Bond" targets alloy phase formation and bond characteristics using established descriptors [31–35]. Mean atomic packing efficiency (Mean APE) was computed with Miracle radii (Matminer-consistent); all other radius-based terms used metallic radii at coordination 12, consistent with crystalline solid solutions and Magnone et al. [32].

Table 1. Nominal and actual split sizes for the training set and the held-out test set.

| Nominal size | Actual size | Train (unique \| samples) | Test (unique \| samples) |
|---|---|---|---|
| **80/20** | 80/20 | 258 \| 1788 | 70 \| 435 |
| **85/15** | 85/15 | 276 \| 1883 | 52 \| 340 |
| **90/10** | ~95/5 | 299 \| 2092 | 29 \| 131 |

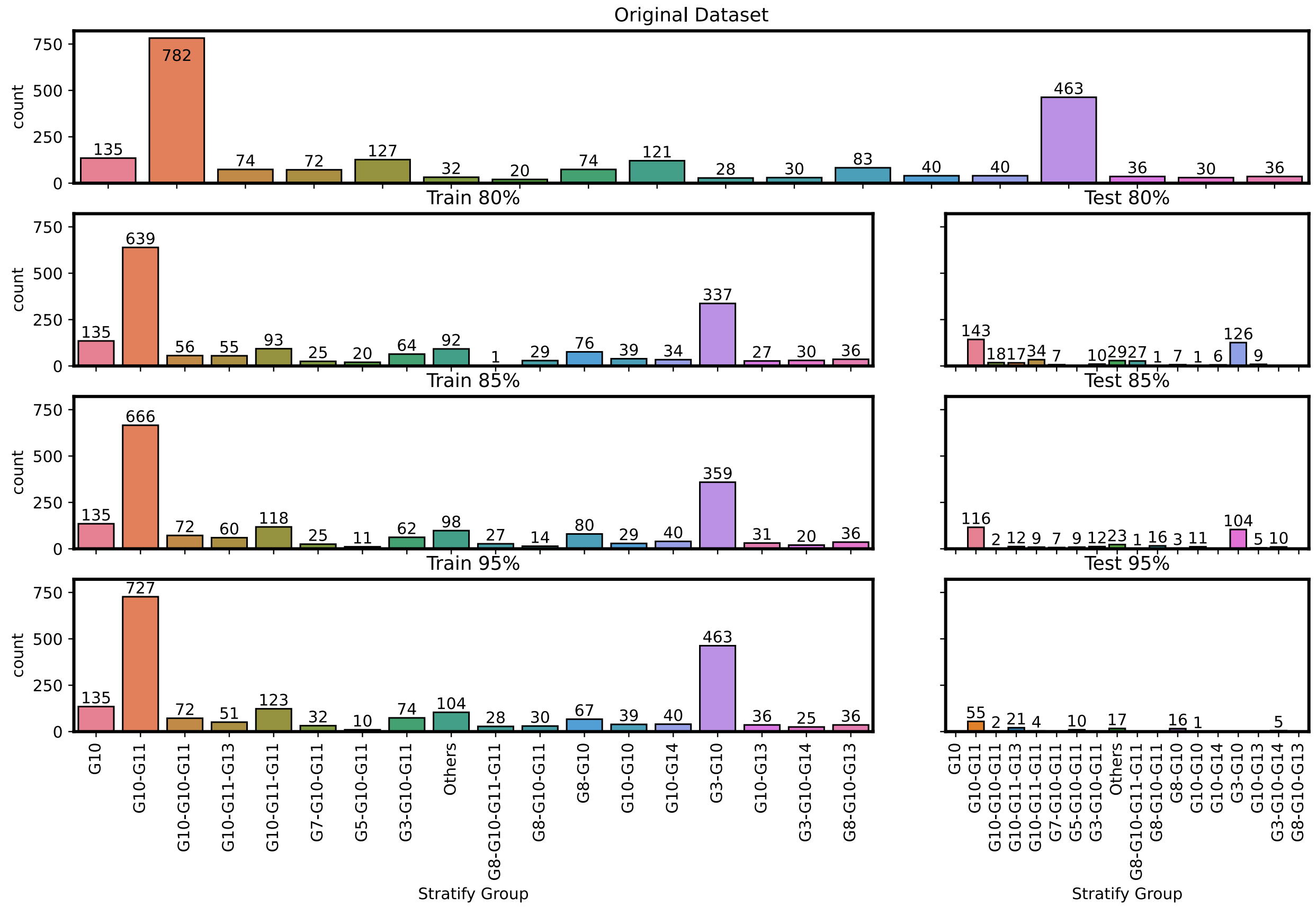

Fig. 1. Counts by stratum in the original dataset and across 80/20, 85/15, ~95/5 training and held-out test splits.

"CBFV" represents composition via rule-of-mixtures properties given by:

$$Property_{alloy} = \sum_{i=1}^{n} c_i P_i \quad (3)$$

where elemental fractions $c_i$ and elemental property tables $P_i$ taken from the 2024 Oliynyk dataset [36]. "Elemental" encodes elemental fractions, allowing the model to learn nonlinear interactions directly from stoichiometry. The full features list appears in Table S1 (Supplementary Material). We formed pairwise, triplet, and quartet combinations to probe complementarity between physics-informed and composition-only information: Exp_Bond, Exp_CBFV, Exp_Elemental, Exp_Bond_CBFV, Exp_Bond_Elemental, Exp_CBFV_Elemental, and Exp_Bond_CBFV_Elemental. Table 2 reports descriptor counts per set.

Table 2. Type and cardinality of primary feature sets

| Feature set | Number of features |
|---|---|
| **Exp** | 4 |
| **Bond** | 18 |
| **CBFV** | 75 |
| **Elemental** | 40 |

This ablation design quantifies the incremental value of experimental context, bond and phase-formation descriptors, and composition-based vectors for permeability prediction.

## 2.5. Model development and validation

Our small and sparse data regime necessitates a shallow ML model that can learn non-linear relationships. This motivated the use of gradient boosting with symmetric ("oblivious") decision trees as implemented in CatBoost to model the non-linear relationships between descriptors and permeability. CatBoost was selected because: (1) ordered boosting reduces prediction shift and target leakage; (2) symmetric depth-limited trees provide strong built-in regularization on small tabular datasets; (3) the library offers native handling of missing values and robust training controls (learning-rate shrinkage, L2 regularization, subsampling, early stopping) to limit overfitting [37]; (4) tree ensembles are comparatively insensitive to feature collinearity; (5) and CatBoost's regularization further mitigates variance when many descriptors are included.

### 2.5.1 Modelling details

We examined two modeling strategies: one using all features and another with feature selection, to identify robust models that balance performance, complexity, and extrapolative capability. Throughout this work, model performance was assessed using mean absolute error (MAE), root mean squared error (RMSE), and coefficient of determination $R^2$ (see the "4. Model Evaluation" section in the Supplementary Material for details). In addition, the source code to reproduce our work and visualize the results is freely available in GitHub (https://github.com/AldoKwamibar/Pd-membranes-permeability).

In the first strategy, CatBoost regressors were trained directly on each raw feature set (see Section 2.4) to assess the model's robustness to multicollinearity and high-dimensional input spaces. Model hyperparameters were optimized through randomized search using a stratified group K-fold cross-validation scheme with $k \in \{3, 5, 10\}$ to accommodate unequal group sizes. The folds were stratified based on the periodic group similarity encoding described earlier to preserve composition-family distributions, and mutual exclusivity of formula groups across folds was ensured. Early stopping was applied as a regularization technique to prevent overfitting, halting training when the validation error ceased to improve for 20 consecutive iterations. During hyperparameter optimization, the same held-out validation fold was used as internal evaluation set for both tracking the loss function and early stopping, without data leakage [38]. The initial number of estimators was set to 10,000 as an upper limit with shrinkage, while the effective model size was determined automatically by early stopping. Finally, after optimization, the model was retrained on the complete training set using the selected hyperparameters and the median optimal number of estimators identified during early stopping. The external test set was kept untouched until final evaluation.

In the second strategy, a two-stage feature selection scheme was applied to reduce multicollinearity and dimensionality, producing more compact and interpretable feature sets. In the first stage, feature redundancy was analyzed by calculating pairwise Pearson correlation coefficients. A threshold of $|r| > 0.90$ was used to identify strongly correlated pairs. When two features met this criterion, one was retained based on domain knowledge, prioritizing properties historically associated with hydrogen permeation in metals. Detailed information regarding feature filtering is provided in the "5. Pearson Correlation Filtering" section of the Supplementary Material. After reducing multicollinearity using the $|r| > 0.90$ filter, dimensionality was further decreased through a two-step procedure: A fold-optimized recursive feature elimination (RFE) guided by SHapley Additive Explanations (SHAP) values, followed by union aggregation of the top-ranked features across folds. The RFE was implemented using CatBoost's built-in *RecursiveBySHAP* algorithm and applied independently to each fold to determine the optimal number of important features identified by the SHAP framework. SHAP quantifies the contribution of each descriptor to model predictions, providing a consistent feature importance measure [39]. After RFE, a tree-specific SHAP analysis (*TreeSHAP*) was recomputed for each fold to rank descriptors from most to least influential.

For each target subset size, $m\epsilon\{10,12,15,17,20,22,25,30\}$, the top $m$ features were selected within each fold, and their union was taken across folds to form an aggregated feature set for that $m$. Because fold-specific lists overlapped only partially, the aggregated set size was typically close to $m$. Each aggregated feature set was then used to train CatBoost models with default hyperparameters for comparison. During training, early stopping with cross-validation (patience = 20) was applied to determine the optimal number of estimators per fold. The median value of these optimal estimators was subsequently used to retrain the model on the full training partition. The external test set remained untouched until the final prediction stage.

#### 2.5.2. Selection of final parsimonious model

Each aggregated feature set produced a separate CatBoost regressor, resulting in multiple models that differed in feature set composition and number of estimators. Test predictions were evaluated using block bootstrap to estimate uncertainty in the performance metrics. Details on block bootstrap are available in section '10. Confidence interval' of the Supplementary Material. The one-standard-error rule was then applied to select the most parsimonious model whose mean absolute error (MAE) was within one standard error of the minimum MAE [40].

### 2.6. Virtual screening of B2-phase stabilized Pd–Cu–M ternary alloys

We used the simplest-and-most-efficient model (see sec. 2.5.2.) to construct composition–permeability contour maps at 673 K and 131.0 $Pa^{0.5}$ for 16 metallic elements, potential B2-phase stabilizers. The virtual alloys follow the general formula $Pd_{(100-x-y)}Cu_xM_y$ assuming 15 µm thickness. The compositions satisfied the constraints 35 ≤ $100 - x - y$ ≤ 50 and 0 ≤ $y$ ≤ 20, with a resolution of 0.1%, where M ∈ {Al, Cr, Mn, Fe, Ga, Hf, La, Mg, Nb, Sc, Ta,

Ti, V, Y, Zn, Zr} were selected based on previous DFT works [14,15].

## 3. Results and discussions

### 3.1. Dataset for modeling

After preprocessing, the final dataset contains 2,223 records spanning 328 unique canonical compositions. By number of components, the set includes pure Pd, 186 binaries, 131 ternaries, 7 quaternaries, 0 quinary, 1 senary, and 2 septenary alloys. Thickness ranges from 1 μm to 1100 μm, temperature from ≈292 K to ≈1175 K, and permeability from 0 to $9.54\times10^{-8}$ $mol\cdot m^{-1}\cdot s^{-1}\cdot Pa^{-0.5}$. Details of individual parameter distributions are provided in Fig. S1 (Supplementary Material). A dominant part of the dataset contains Pd-rich substitutionally random *fcc* solid solutions, while B2 and mixed *fcc*/B2 alloys were always Cu-containing alloys. The elemental space covers 40 elements including transition metals (3d–5d), a metalloid (B), selected post-transition metals, and lanthanides. Fig. 2 depicts a pictorial summary: panel A maps the elemental coverage, and panel B presents the distribution of Bravais lattice types based on column-similarity grouping.

### 3.2. Column similarity encoding and stratified data splitting for regression

The mapping analysis (Fig. 2a and Fig. 2b) shows that the dataset is strongly imbalanced: four transition metals dominate (Cu >> Ag > Au > Y), with other elements sparsely represented. Lower-component alloys are overrepresented (binaries and ternaries >> higher-order systems), and *fcc* structures greatly outnumbered other Bravais lattice types. Copper alone accounts for 57.1% of the entries, reflecting both the search for low-cost, sulfur-resistant membranes and the incremental adjustments traditionally applied to optimize Pd–Cu compositions and operating conditions. Such redundancy and imbalance are common in materials datasets, and a model trained directly on them risks misleading performance estimates, particularly for out-of-distribution tasks such as predicting underrepresented systems or alloys containing elements outside the current chemical space. Similar concerns about inflated accuracy from random splits in redundant inorganic datasets have been raised in prior works [42,43].

To mitigate these biases, we introduced a chemistry-aware partitioning scheme (Sec. 2.3). Elements were encoded by their column position in the periodic table, and the so obtained data were stratified to enforce approximate uniformity of encoded families across training, validation, and test sets.

This design encourages the model to transfer learning across systems with similar valence-electron chemistry, improving robustness to underrepresented alloys. For example, as shown in Fig. 2a, Pd–(Cu, Ag, Au) alloys can be grouped as G10–G11, and Pd–(Y, lanthanides) alloys as G3–G10, thereby partially rebalancing *fcc*-rich families and forcing the model to treat them as comparable chemical groups. This strategy reduces covariate shift by maintaining similar encoded distributions across partitions, where covariate shift refers to cases in which the distributions of training and test inputs differ while the conditional output distribution remains unchanged [44].

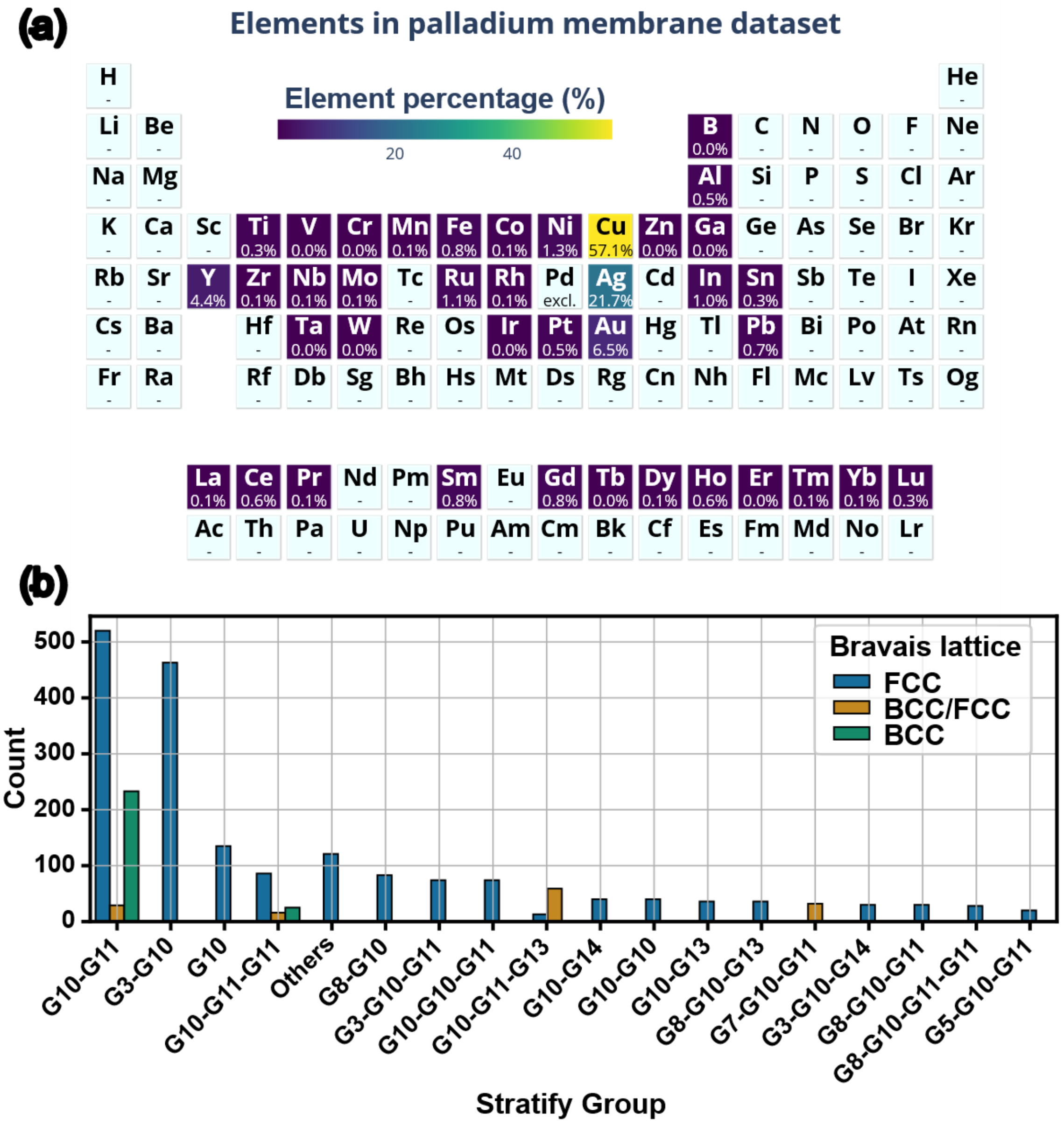

Fig. 2. (a) Elemental composition in the palladium alloy membrane dataset (plot generated using pymatviz [41]); percentages exclude Pd, the main element. Cells with 0.0% indicate values < 0.1%. (b) Bravais lattice type by stratum.

Although residual shifts may persist in non-compositional variables such as thickness, the column-similarity encoding provides a more chemically meaningful and balanced split than random or purely stratified schemes. A systematic benchmarking against other material-specific splitting methods is beyond the scope of this work, but will be valuable in future studies.

### 3.3. Model performance without feature selection

In materials property prediction, two strategies are commonly adopted: a model-centric approach, which compares the performance of multiple algorithms with minimal preprocessing, and a feature-centric approach, which emphasizes advanced feature engineering with a single model to maximize predictive accuracy. We adopted the latter, selecting CatBoost regression under the heuristic that predictive power is primarily locked within the representation of the data rather than the model itself. This is consistent with the "no-free-lunch" theorem, which states that no algorithm universally outperforms others across all problems, and that accuracy can be improved through appropriate feature design [45]. Recent studies in materials informatics have also demonstrated the benefit of carefully crafted descriptors over brute-force model comparisons [46,47].

After hyperparameter tuning, the performance of the descriptor sets was compared across 80/20, 85/15, and ~ 95/5 train/test splits under identical stratified group K-fold cross-validation ($k \in \{3, 5, 10\}$). The performance metrics employed are complementary: MAE ($mol \cdot m^{-1} \cdot s^{-1} \cdot Pa^{-0.5}$) is the average absolute distance between the actual and predicted values and is less sensitive to extreme values; RMSE ($mol \cdot m^{-1} \cdot s^{-1} \cdot Pa^{-0.5}$) emphasizes large errors and is preferred when outliers are exponentially rare; $R^2$ (unitless) quantifies explained variance. For a physically sound permeability prediction, both MAE and RMSE should be comparable or lower than the experimental uncertainties, which cannot be readily obtained for our multi-source dataset.

We arranged the results obtained at this stage in Fig. 3 ((a)–(f)), where prediction results on the unseen test set are vertically stacked on the right side and those on the training set on the left. Training $R^2$ values ranged from 0.79 to 0.99, indicating apparently good fitting capacity. However, the 95/5 split produced deceptively low MAE and RMSE but consistently negative $R^2$ on the held-out test set. This is caused by the very small test partition, limited target variance and the fact that the compositions in the 5% test set are rare and do not cover a representative part of the compositional distribution of the 95% training set. By contrast, the 80/20 split yielded the highest overall errors, reflecting an unfavorable bias–variance trade-off. The 85/15 split provided the most reliable balance: sufficient training size, a large enough test block for stable estimates, and good distributional concordance across column-similarity strata. We therefore adopted 85/15 for subsequent analyses. Negative $R^2$ in the 95/5 case is not anomalous but reflects the definition used in the scikit-learn ML library [48], where a small variance in test target can drive $R^2$ below zero. Diagnostics confirm this: the 95/5 test set has a three-times smaller standard deviation and ~1.4 orders of magnitude lower total sum of squares than the 85/15 test set (see Supplementary Material, Section 9). From the aforementioned results, the analyses using 80/20 and 95/15 splits were considered as diagnostic tests rather than definitive models.

Within the 85/15 split, no consistent dependence was observed between scoring metrics and the number of *k*-folds. Importantly, combined MAE and RMSE values fall within the interquartile range of the full permeability distribution ([5.90×10$^{-9}$, 2.57×10$^{-8}$ mol·m$^{-1}$·s$^{-1}$·Pa$^{-0.5}$]), confirming that the models capture the correct magnitude of the property space. However, prediction quality remains unsatisfactory for very low-permeability alloys such as Pd–Fe and Cu-rich *fcc* Pd–Cu [28,30], which lie in the skewed distribution tail and are likely treated as outliers. This is reflected in RMSE values being an order of magnitude larger than MAE, consistent with the higher sensitivity of RMSE to outliers. For context, most permeability values in the dataset range from 10$^{-9}$–10$^{-8}$ mol·m$^{-1}$·s$^{-1}$·Pa$^{-0.5}$, while peak values for Pd-based membranes (composite or self-standing) typically lie near 10$^{-8}$ [49].

Among the descriptor families, the simplest encoding (Exp_Elemental) achieved the strongest held-out performance ($R^2$ = 0.80, MAE = 7.09×10$^{-9}$ mol·m$^{-1}$·s$^{-1}$·Pa$^{-0.5}$, RMSE = 0.97×10$^{-8}$ mol·m$^{-1}$·s$^{-1}$·Pa$^{-0.5}$, based on 5-fold CV). This is surprising, given fractional stoichiometry representations are the simplest and are often regarded as more efficient only for larger datasets, as shown in the ElemNet deep-learning model [50,51]. In contrast, the domain-grounded Exp_Bond set (22 descriptors) delivered stable performance across each CV folds [*k*=3 ( $R^2_{train}$ =0.91, $R^2_{test}$ =0.78), *k*=5 (0.87, 0.77), *k*=10 (0.86, 0.76)], underscoring the robustness of alloy phase-formation descriptors. These findings align with Magnone et al. (2023), who emphasized the predictive value of valence electron concentration (VEC), electronegativity difference, mixing entropy, and atomic radius difference, and with Yan et al. (2025), who applied similar principles to design Nb-based membranes [32,34].

Overall, Fig. 3 highlighted a trade-off: Exp_Elemental provides the highest raw accuracy but limited extrapolative power, while Exp_Bond offers consistent, interpretable predictions grounded in alloy chemistry. The Exp_Bond_Elemental combination strikes a practical middle ground, capturing the benefits of both. These results justify our focus on domain-knowledge-containing descriptor sets for subsequent feature selection and screening analyses.

### 3.4. Dimensionality–complexity reduction

After identifying the 85/15 split as the most stable partition (Sec. 3.3), we next sought to balance accuracy and model simplicity. Each feature set was first filtered for redundancy ($|r| \geq 0.90$) and then subjected to fold-wise SHAP-based recursive feature elimination with union aggregation across folds. This procedure yielded 135 models differing in cross-validation scheme, feature pool, and subset size, enabling a systematic assessment of the performance–complexity trade-off.

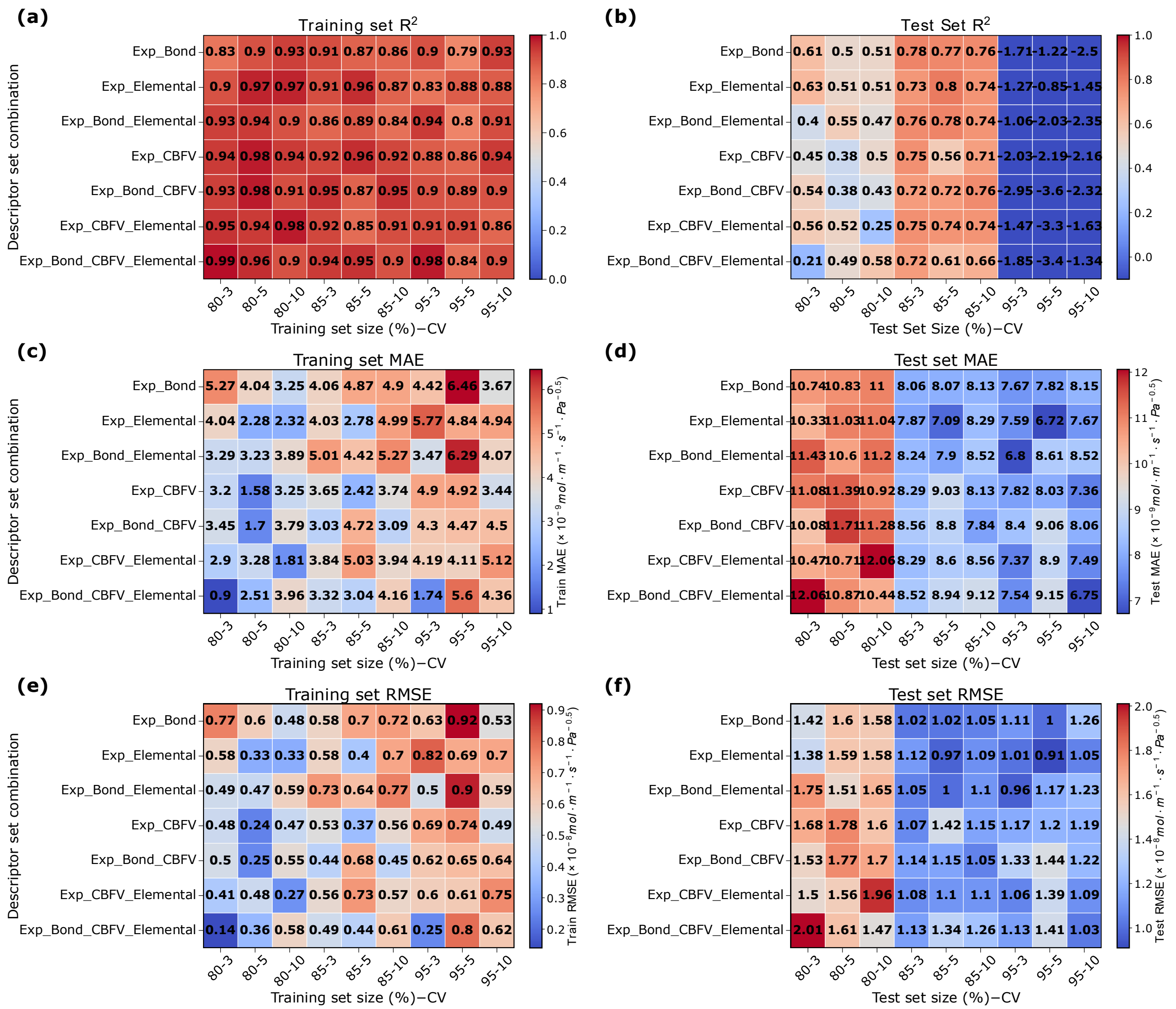

Fig. 3. (a) – (f) Fitting and prediction performance on the training and held-out test sets (without feature selection). The left panels show training results; the right panels show test results. For each descriptor set (*y*-axis), a cell corresponds to a specific train/test split ratio and cross-validation configuration (*x*-axis). In each grid, the descriptor sets are arranged from the lowest cardinality set to the highest (top-to-bottom).

Fig. 4 shows the evolution of $R^2$ on the held-out test set as a function of feature count. It can be seen that parsimonious sets consistently achieved higher $R^2$ than larger ones, underscoring the value of compact, physics-informed descriptors. The best models reached $R^2_{test}$ = 0.82, including: (*i*) 38 features from the Exp_Bond_CBFV_Elemental pool (3-fold CV), (*ii*) 15 features from Exp_Bond (3-fold CV, $R^2_{train,\ mean}$ = 0.94), and (*iii*) 19 features from Exp_Bond_CBFV (3-fold CV). Relative to predictions without feature selection, these models showed reduced overfitting and slightly improved generalization, while requiring fewer features. Notably, the 15-feature Exp_Bond set delivered equivalent test performance to larger sets, highlighting the sufficiency of alloy phase-formation descriptors. Our selected subset includes 15 of the 20 alloy chemistry parameters identified by Wen et al. (2019) [33], extending beyond those emphasized by Magnone et al. [32,33].

It should be emphasized that although models using elemental fractions as direct inputs performed strongly, they can struggle to extrapolate to alloys containing elements outside the training space. Such models are therefore better suited for interpolation (prediction for an element in the same elemental space) within known systems rather than out-of-distribution prediction. Overall, these results demonstrate that rigorous feature selection enables reduction of model complexity without loss of performance, and even modest gains, re-emphasizing the use of compact domain-informed descriptors for virtual screening.

### 3.5. Final model selection

To select a final model for virtual screening, we prioritized parsimony and generalization. Model selection was guided by Occam's razor and the one-standard-error (one-SE) rule, which recommend choosing the simplest model whose accuracy is statistically indistinguishable from the best [52–54]. Performance uncertainty was quantified by block bootstrapping of test-set predictions ($n$ = 50,000 resamples, grouped by formula), yielding mean scores, standard errors, and 95% confidence intervals.

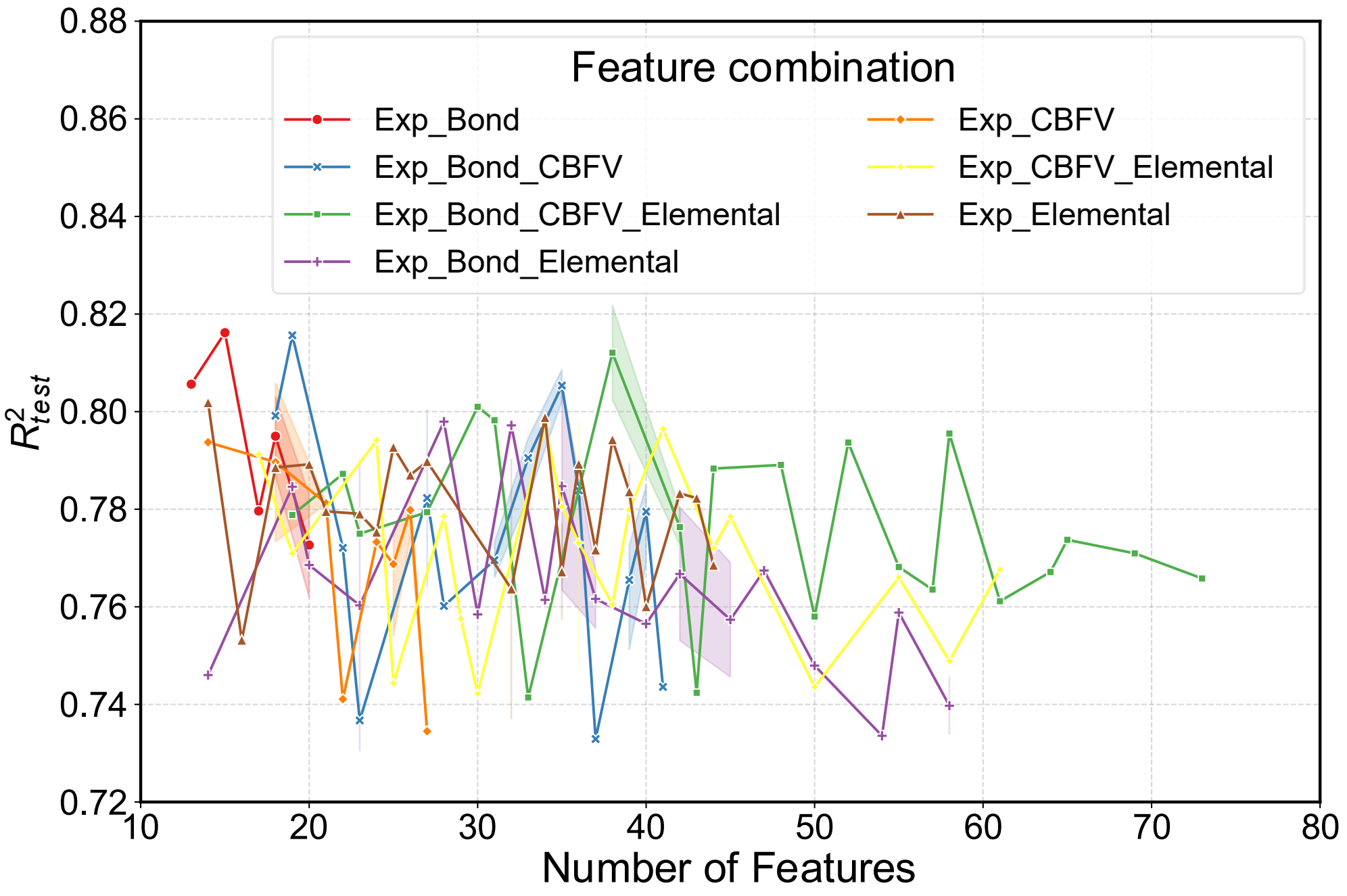

Fig. 4. Coefficient of determination ($R^2$) on the held-out test set as a function of the number of features.

As shown in Fig. 5, the one-SE criterion identified the Exp_Bond model with 13 features as the most parsimonious and efficient. This domain-informed model achieved MAE = $7.4\times10^{-9}$ mol·m$^{-1}$·s$^{-1}$·Pa$^{-0.5}$ and RMSE = $9.5\times10^{-9}$ mol·m$^{-1}$·s$^{-1}$·Pa$^{-0.5}$, values comparable to the lowest errors from larger feature sets but with far fewer descriptors. Notably, RMSE is now in the same order of magnitude as MAE, indicating improved robustness to outliers.

The parity plot in Fig. 6 (point predictions) shows that the model explains 81% of the variance in $H_2$ permeability ($R^2$ = 0.81) across diverse alloy groups. Bootstrap resampling confirmed this performance with a mean $R^2$ = 0.78 and a 95% confidence interval [0.61, 0.88] (Table 3). Note that the difference in the results of block bootstrap (Table 3) and point predictions (Fig. 6) is expected since block bootstrap is a stochastic method, corresponding to a resampling with replacement of groups of formulas and associated prediction from the test set, followed by scoring multiple times. The asymmetric interval, with greater uncertainty on the lower side, reflects the model's reduced confidence for very low-permeability alloys, which act as statistical outliers. This highlights an opportunity: expanding the dataset with additional low-permeability alloys would likely improve calibration and reduce skewness in future iterations. Overall, the selected Exp_Bond model balances accuracy, interpretability, and parsimony, making it a robust tool for virtual screening of Pd-based alloys.

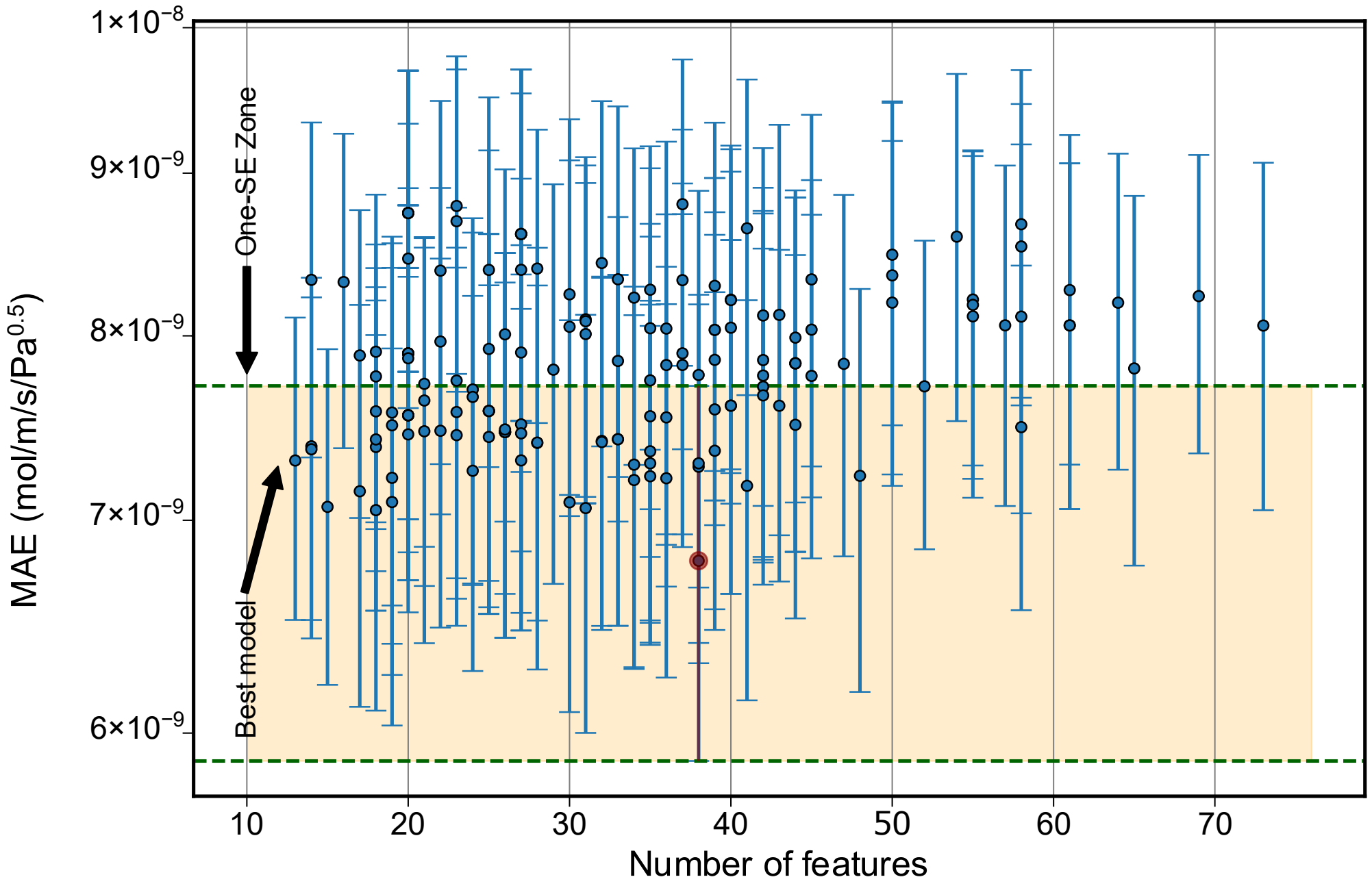

Fig. 5. Mean MAE (o-shaped markers with black edge color) with standard errors; the shaded band from the mean denotes the one-SE region. Error bars represent standard errors obtained using bootstrap. The model achieving the lowest MAE, highest $R^2$, and lowest RMSE is highlighted in red, and the selected model is indicated by a pointing black arrow.

Table 3: Results of block bootstrap on the test set predictions by using the most parsimonious model trained on Exp_Bond (13 features). The mean, standard error, the [2.5, 97.5] percentiles to compute the 95% CI ([lower, upper]) bounds are reported. The results are obtained after $n$ = 50,000 resampling with replacement. The model is trained once and remained fixed.

| Test metrics | Mean | Std. error | Lower bound | Upper bound |
|---|---|---|---|---|
| $R^2$ | 0.78 | 0.070 | 0.614 | 0.876 |
| **MAE** | $7.31\times10^{-9}$ | $7.97\times10^{-10}$ | $5.68\times10^{-9}$ | $8.77\times10^{-9}$ |
| **RMSE** | $9.37\times10^{-9}$ | $8.75\times10^{-10}$ | $7.48\times10^{-9}$ | $1.09\times10^{-8}$ |

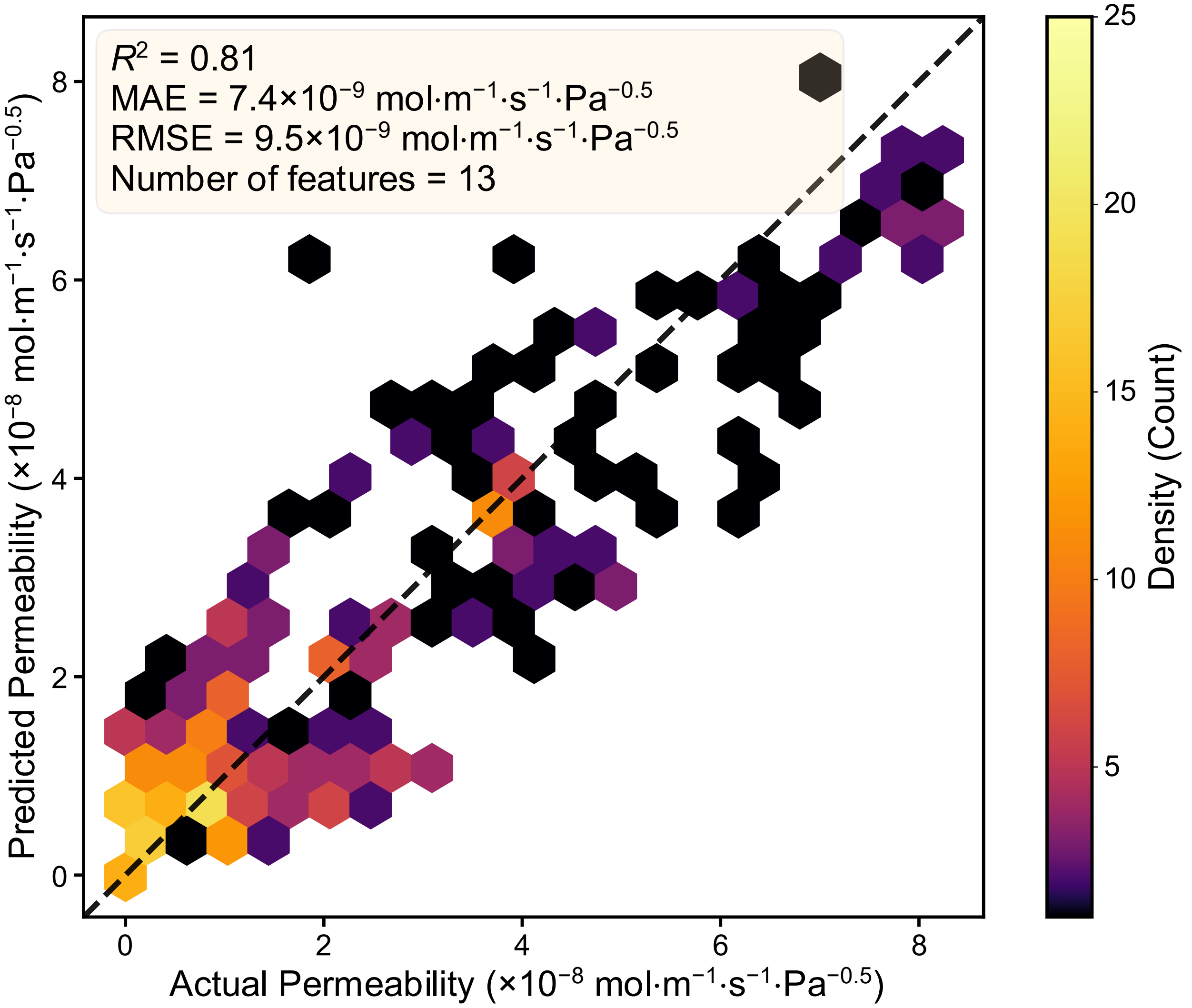

Fig. 6. Parity plot of point predictions on the held-out test set for the 13-feature Exp_Bond model selected by the one-SE rule.

### 3.6. SHAP for feature importance analysis

We applied TreeSHAP to the final 13-feature Exp_Bond model to identify the descriptors most strongly influencing hydrogen permeability (Fig. 7). The normalized lattice size difference relative to pure Pd ($\Delta a_{ss}/a_{Pd}$) emerged as the dominant factor (Fig. 7a). Larger lattice expansion generally increased predicted permeability (Fig. 7b), consistent with established behavior in *fcc* Pd-rich solid solutions [55–58,58]. Expansion makes $H_2$ dissolution more exothermic, improving the solubility–diffusivity balance, whereas contraction enhances diffusivity but reduces solubility, leading to modest decreases in permeability. This explains why Pd–rare-earth solid solutions, with significantly larger solute radii, often outperform Pd–noble metal alloys.

Environmental variables were accurately highlighted by the TreeSHAP analysis. The pressure difference $\Delta P^{0.5}$ and thickness appeared apparently influential, although permeability is theoretically an intrinsic material property and independent of these terms (Gallucci, 2012; Kolor et al., 2024) [1]. This can be accessed by considering the full Richardson equation ( $J = \frac{\phi}{l}\,\Delta P^{0.5}$ ). The permeance instead has an inverse relationship with the thickness when the diffusion is bulk-controlled [59]. Their apparent importance likely reflects systematic errors introduced during semi-automatic data extraction and noise introduced by pointwise recomputation of permeability from $J - \Delta P^n$ data, which TreeSHAP correctly detected. This apparent discrepancy reflects the small but systematic gap between theory and experiment: pointwise permeability, recomputed from digitized $J - \Delta P^n$ curves, inherits residual dependence on $\Delta P^n$ and thickness. The model correctly captured this artefact, highlighting its sensitivity to experimental noise and data processing rather than a violation of Sieverts' law. In contrast, temperature displayed a physically meaningful effect: permeability jointly depends on the diffusivity (D) and solubility (S) which increase with the temperature, in line with Arrhenius behavior ($\phi = D \cdot S = \phi_0 \exp\left(\frac{-E_a}{RT}\right)$).

Among alloy descriptors, the atomic size difference (δ) showed a two-regime effect: medium-to-large δ values enhanced permeability, while small δ reduced it. The melting temperature ( $T_m$ ) also correlated positively with permeability in many cases, though not universally. For Group 11 alloys (Cu, Ag, Au), higher $T_m$ does not correspond to higher permeability ( $\phi_{Pd-Ag} > \phi_{Pd-Au} \approx \phi_{Pd-Cu}$). However, when comparing Pd–Ag and Pd–Y, the correlation reemerges. We suggest that higher $T_m$ raises Tammann (~$0.5T_m$) and Hüttig (~$0.3T_m$) temperatures, delaying bulk atomic mobility and surface segregation that degrade $H_2$ transport. Thus, while $T_m$ is not a universal predictor, it captures bond-strengthening effects in certain systems.

Thermodynamic descriptors also showed systematic influence: higher mixing enthalpy and entropy were associated with reduced permeability, reflecting the destabilization of solid solutions. This highlights solid-solution stability as a key prerequisite for high hydrogen transport. Interestingly, the VEC, previously emphasized by Magnone et al. [32], was not ranked decisive in this study, likely because tree-based models prioritize nonlinear combinations of descriptors, which can diminish the marginal role of VEC. Instead, TreeSHAP identified secondary features such as the mixing entropy-to-squared size ratio (Λ), modulus mismatch in the strengthening model (η), mismatch of local electronegativity (D·χ), and shear modulus difference (δG) as contributing factors, suggesting that permeability depends on a broader interplay of structural, thermodynamic, and electronic descriptors rather than a single dominant parameter.

Taken together, these results show that the ML model captures both expected physical trends (lattice expansion, Arrhenius temperature dependence, stability penalties) and subtler nonlinear effects. The analysis provides mechanistic insight and highlights which alloy properties most consistently favor enhanced permeability.

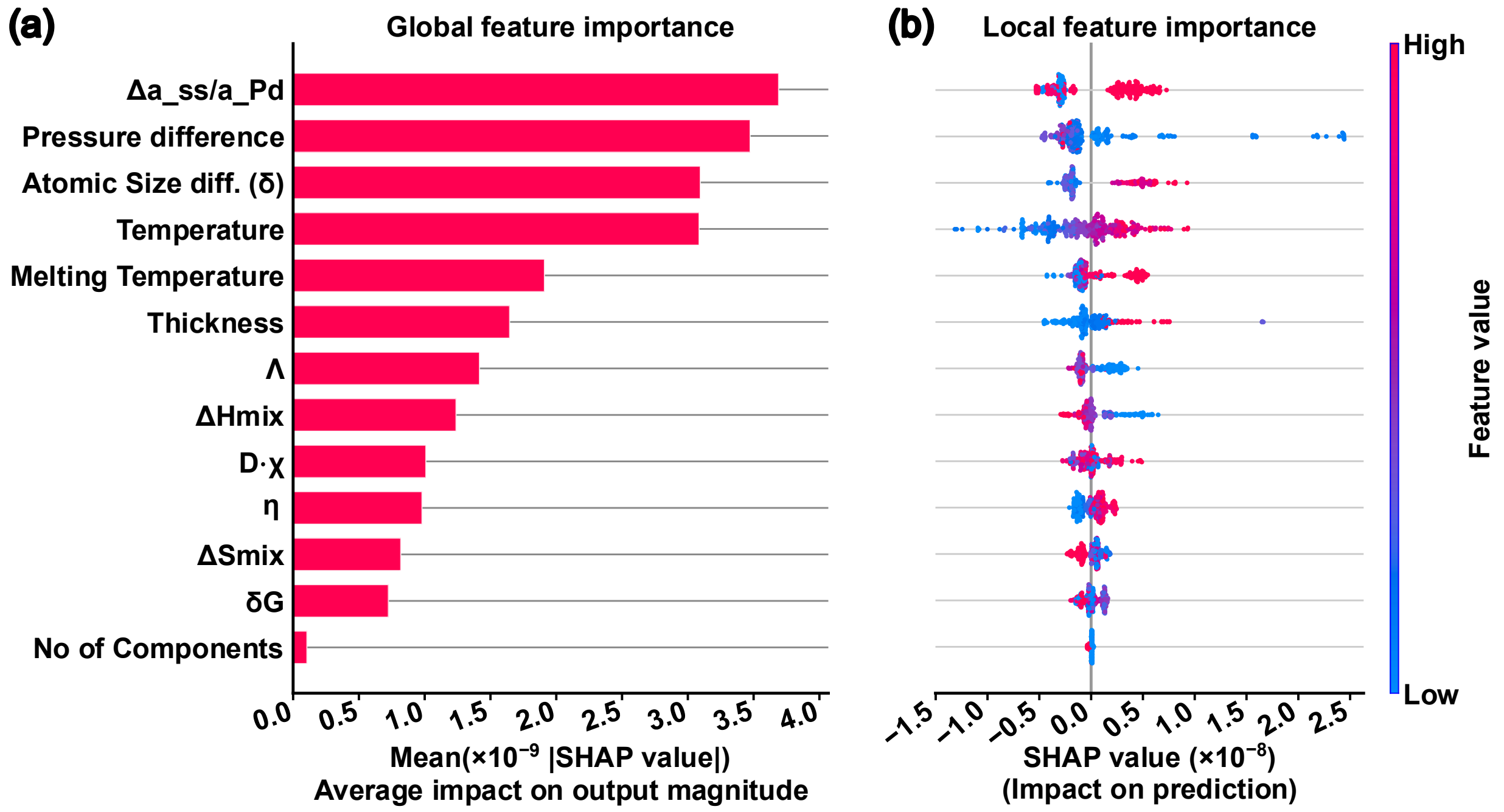

Fig. 7. (a) Global feature importance (bar plot), and (b) Local feature importance (bee swarm plot) of the simplest–efficient 13-feature Exp_Bond model.

### 3.7. Mapping $H_2$ permeability of potential B2-phase stabilized $Pd_{(100-x-y)}Cu_xM_y$

We used the final 13-feature Exp_Bond model to construct composition–permeability contour maps for 16 ternary systems of the form $Pd_{(100-x-y)}Cu_xM_y$ ( $35 \le 100 - x - y \le 50$, $0.1 \le y \le 20$), corresponding to Pd substitution by M on the B2–PdCu sublattice. The scan, performed at 0.1 at% resolution, covered 453,000 virtual alloys. The chosen stabilizers M were drawn from first-principles studies identifying elements that promote negative formation enthalpies in B2 Pd–Cu–M ternaries [14,15]. The search space respects the known B2 stability window of PdCu (~36–47 at% Pd) [4], with a focus on Pd-lean regions to reduce cost.

Fig. 8 reveals several alloy families with predicted permeabilities exceeding the benchmark B2− $Pd_{60}Cu_{40}$ ($\phi > 3.10^{-8}$ mol·m$^{-1}$·s$^{-1}$·Pa$^{-0.5}$ [2]). Rare-earth and group IV solutes (Y, Sc, La, Zr, Hf), together with Mg, consistently achieved maxima above $3\times10^{-8}$ mol·m$^{-1}$·s$^{-1}$·Pa$^{-0.5}$, nearly double the reference. These include elements absent from the training dataset (Mg, Sc, Hf, Zn), representing true extrapolative (out-of-distribution) predictions.

To avoid bias toward permeability-only extremes, we constructed Pareto fronts balancing three criteria: maximum predicted permeability, minimum Pd content, and lowest Miedema formation enthalpy [60]. Examples are summarized in Table 4. While several compositions approach twice the Pd–Cu benchmark, alloys with very low Pd fractions may sacrifice catalytic surface density, and some solutes (e.g., rare earths) exhibit miscibility gaps at low concentrations [61]. We therefore applied two additional filters. First, surface segregation energies from Ruban et al. [62], indicate that Ti, V, Cr, Fe, Zr, Nb, Hf, and Ta are anti-segregating (positive segregation energy), making them less likely to poison Pd active sites. Second, analysis of Pd–M binary phase diagrams [63] identified concentration windows where M forms a complete solid solution with Pd in the $Fm\bar{3}m$ structure.

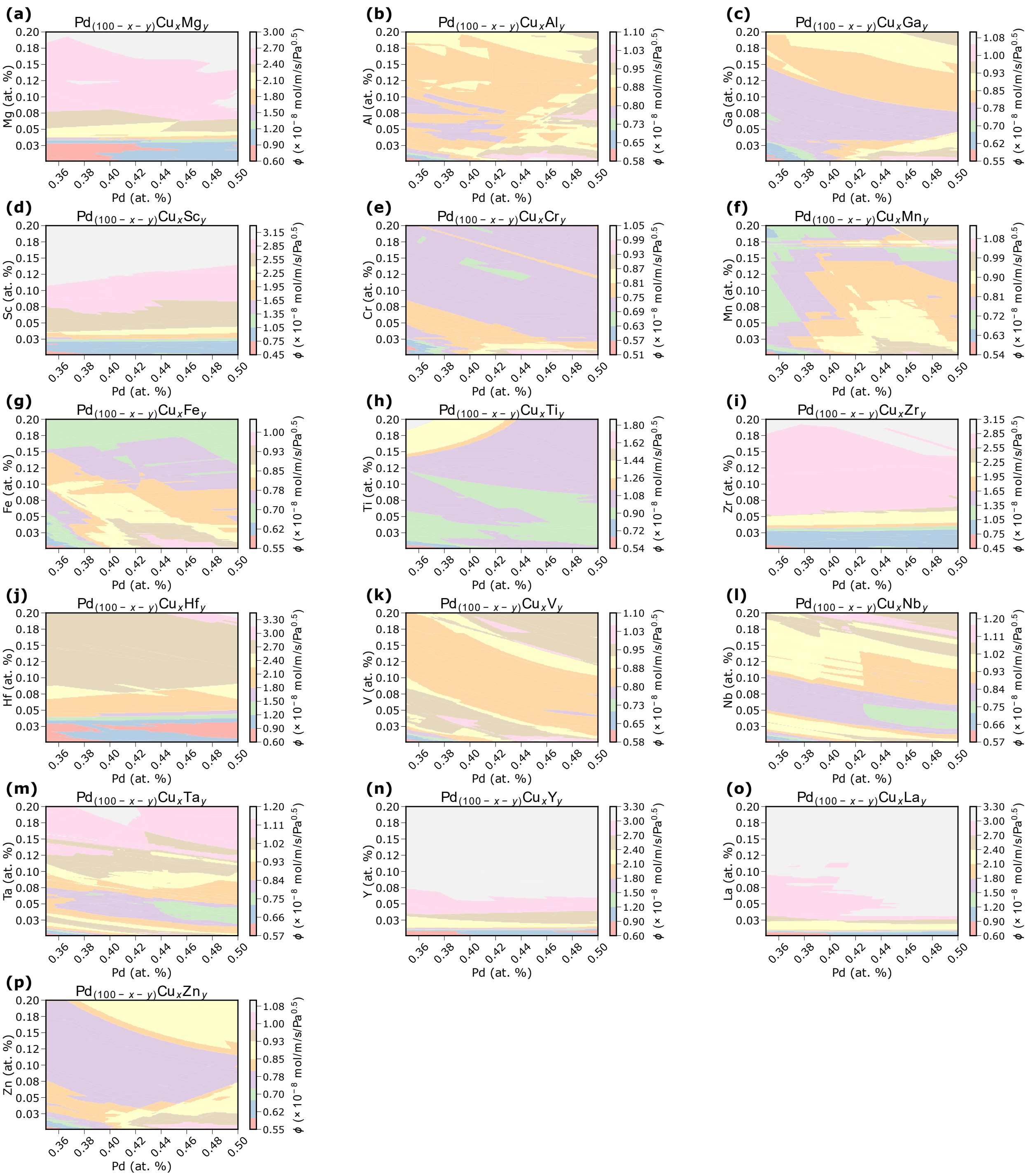

Fig. 8. (a) – (p) Predictive $H_2$ permeability at 673.15 K, 131.0 $Pa^{0.5}$ for 15 μm-thick 16 virtual ternary $Pd_{(100-x-y)}Cu_xM_y$ alloy systems based on the 13-feature Exp_Bond model. (a) – (p) Ternary system of $Pd_{(100-x-y)}Cu_xM_y$ (M∈{Mg, Al, Ga, Sc, Cr, Mn, Fe, Ti, Zr, Hf, V, Nb, Ta, Y, La, Zn}).

Table 4. Pareto set of extracted compositions (in atom% and weight%) with the highest predicted permeability (max. $\phi$), minimum Pd content (min. Pd) and minimum enthalpy (min $\Delta H_{mied}$) for each system. Highlighted formulas (bold) are the composition satisfying the miscibility gap filter.

| Virtual system | | Virtual candidates (at %) | Virtual candidates (wt %) | Permeability (mol·m$^{-1}$·s$^{-1}$·Pa$^{-0.5}$) |
|---|---|---|---|---|
| Mg | max. $\phi$ | $Pd_{35.9}Cu_{44.1}Mg_{20.0}$ | $Pd_{53.74}Cu_{39.42}Mg_{6.84}$ | $2.89\times10^{-8}$ |
| | min Pd | $Pd_{35.0}Cu_{45.0}Mg_{20.0}$ | **$Pd_{52.68}Cu_{40.44}Mg_{6.88}$** | $2.88\times10^{-8}$ |
| | min $\Delta H_{mied}$ | $Pd_{48.0}Cu_{32.0}Mg_{20.0}$ | $Pd_{66.97}Cu_{26.66}Mg_{6.37}$ | $2.80\times10^{-8}$ |
| Al | max. $\phi$ | $Pd_{47.0}Cu_{52.9}Al_{0.1}$ | **$Pd_{59.79}Cu_{40.18}Al_{0.03}$** | $1.08\times10^{-8}$ |
| | min Pd | $Pd_{35.0}Cu_{45.0}Al_{20.0}$ | $Pd_{52.29}Cu_{40.14}Al_{7.57}$ | $8.89\times10^{-9}$ |
| | min $\Delta H_{mied}$ | $Pd_{44.3}Cu_{35.7}Al_{20.0}$ | $Pd_{62.67}Cu_{30.16}Al_{7.17}$ | $9.12\times10^{-9}$ |
| Ga | max. $\phi$ | $Pd_{46.9}Cu_{53.0}Ga_{0.1}$ | **$Pd_{59.66}Cu_{40.26}Ga_{0.08}$** | $1.08\times10^{-8}$ |
| | min Pd | $Pd_{35.0}Cu_{47.6}Ga_{17.4}$ | $Pd_{46.77}Cu_{37.99}Ga_{15.24}$ | $8.54\times10^{-9}$ |
| | min $\Delta H_{mied}$ | $Pd_{45.8}Cu_{34.2}Ga_{20.0}$ | $Pd_{57.74}Cu_{25.74}Ga_{16.52}$ | $9.29\times10^{-9}$ |
| Sc | max. $\phi$ | $Pd_{35.0}Cu_{50.7}Sc_{14.3}$ | **$Pd_{49.08}Cu_{42.45}Sc_{8.47}$** | $3.25\times10^{-8}$ |
| | min Pd | $Pd_{35.0}Cu_{50.7}Sc_{14.3}$ | $Pd_{49.08}Cu_{42.45}Sc_{8.47}$ | $3.25\times10^{-8}$ |
| | min $\Delta H_{mied}$ | $Pd_{50.0}Cu_{30.0}Sc_{20.0}$ | $Pd_{65.48}Cu_{23.46}Sc_{11.06}$ | $3.18\times10^{-8}$ |
| Cr | max. $\phi$ | $Pd_{47.6}Cu_{52.3}Cr_{0.1}$ | $Pd_{60.35}Cu_{39.59}Cr_{0.06}$ | $1.05\times10^{-8}$ |
| | min Pd | $Pd_{35.0}Cu_{61.9}Cr_{3.1}$ | **$Pd_{47.63}Cu_{50.31}Cr_{2.06}$** | $7.41\times10^{-9}$ |
| | min $\Delta H_{mied}$ | $Pd_{46.7}Cu_{33.3}Cr_{20.0}$ | $Pd_{61.16}Cu_{26.04}Cr_{12.80}$ | $7.16\times10^{-9}$ |
| Mn | max. $\phi$ | $Pd_{48.6}Cu_{34.0}Mn_{17.4}$ | $Pd_{62.40}Cu_{26.07}Mn_{11.53}$ | $1.12\times10^{-8}$ |
| | min Pd | $Pd_{35.0}Cu_{62.2}Mn_{2.8}$ | $Pd_{47.56}Cu_{50.48}Mn_{1.96}$ | $7.93\times10^{-9}$ |
| | min $\Delta H_{mied}$ | $Pd_{45.1}Cu_{34.9}Mn_{20.0}$ | $Pd_{59.13}Cu_{27.33}Mn_{13.54}$ | $9.66\times10^{-9}$ |
| Fe | max. $\phi$ | $Pd_{48.2}Cu_{51.7}Fe_{0.1}$ | $Pd_{60.91}Cu_{39.02}Fe_{0.07}$ | $1.05\times10^{-8}$ |
| | min Pd | $Pd_{35.0}Cu_{64.9}Fe_{0.1}$ | **$Pd_{47.42}Cu_{52.51}Fe_{0.07}$** | $5.99\times10^{-9}$ |
| | min $\Delta H_{mied}$ | $Pd_{43.4}Cu_{56.5}Fe_{0.1}$ | $Pd_{56.22}Cu_{43.71}Fe_{0.07}$ | $9.15\times10^{-9}$ |
| Ti | max. $\phi$ | $Pd_{35.0}Cu_{45.0}Ti_{20.0}$ | $Pd_{49.39}Cu_{37.92}Ti_{12.69}$ | $1.81\times10^{-8}$ |
| | min Pd | $Pd_{35.0}Cu_{45.0}Ti_{20.0}$ | $Pd_{49.39}Cu_{37.92}Ti_{12.69}$ | $1.81\times10^{-8}$ |
| | min $\Delta H_{mied}$ | $Pd_{50.0}Cu_{30.0}Ti_{20.0}$ | $Pd_{65.01}Cu_{23.29}Ti_{11.70}$ | $1.03\times10^{-8}$ |
| Zr | max. $\phi$ | $Pd_{35.9}Cu_{44.1}Zr_{20.0}$ | $Pd_{45.23}Cu_{33.17}Zr_{21.60}$ | $3.08\times10^{-8}$ |
| | min Pd | $Pd_{35.0}Cu_{45.0}Zr_{20.0}$ | $Pd_{44.29}Cu_{34.01}Zr_{21.70}$ | $3.06\times10^{-8}$ |
| | min $\Delta H_{mied}$ | $Pd_{50.0}Cu_{30.0}Zr_{20.0}$ | $Pd_{58.78}Cu_{21.06}Zr_{20.16}$ | $3.04\times10^{-8}$ |
| Hf | max. $\phi$ | $Pd_{49.6}Cu_{30.4}Hf_{20.0}$ | $Pd_{48.97}Cu_{17.92}Hf_{33.11}$ | $3.42\times10^{-8}$ |
| | min Pd | $Pd_{35.0}Cu_{45.5}Hf_{19.5}$ | $Pd_{36.89}Cu_{28.64}Hf_{34.47}$ | $2.80\times10^{-8}$ |
| | min $\Delta H_{mied}$ | $Pd_{50.0}Cu_{30.0}Hf_{20.0}$ | $Pd_{49.28}Cu_{17.66}Hf_{33.06}$ | $3.41\times10^{-8}$ |
| V | max. $\phi$ | $Pd_{46.0}Cu_{53.9}V_{0.1}$ | **$Pd_{58.80}Cu_{41.14}V_{0.06}$** | $1.09\times10^{-8}$ |
| | min Pd | $Pd_{35.0}Cu_{57.4}V_{7.6}$ | $\mathbf{Pd_{48.00}Cu_{47.01}V_{4.99}}$ | $9.55\times10^{-9}$ |
| | min $\Delta H_{mied}$ | $Pd_{50.0}Cu_{30.0}V_{20.0}$ | $Pd_{64.52}Cu_{23.12}V_{12.36}$ | $9.85\times10^{-9}$ |
| Nb | max. $\phi$ | $Pd_{40.4}Cu_{39.6}Nb_{20.0}$ | $Pd_{49.57}Cu_{29.01}Nb_{21.42}$ | $1.21\times10^{-8}$ |
| | min Pd | $Pd_{35.0}Cu_{45.2}Nb_{19.8}$ | $Pd_{44.15}Cu_{34.05}Nb_{21.80}$ | $1.04\times10^{-8}$ |
| | min $\Delta H_{mied}$ | $Pd_{50.0}Cu_{30.0}Nb_{20.0}$ | $Pd_{58.57}Cu_{20.98}Nb_{20.45}$ | $1.10\times10^{-8}$ |
| Ta | max. $\phi$ | $Pd_{39.5}Cu_{40.6}Ta_{19.9}$ | $Pd_{40.48}Cu_{24.84}Ta_{34.68}$ | $1.20\times10^{-8}$ |
| | min Pd | $Pd_{35.0}Cu_{49.1}Ta_{15.9}$ | $Pd_{38.32}Cu_{32.09}Ta_{29.59}$ | $1.14\times10^{-8}$ |
| | min $\Delta H_{mied}$ | $Pd_{50.0}Cu_{30.0}Ta_{20.0}$ | $Pd_{49.05}Cu_{17.58}Ta_{33.37}$ | $1.13\times10^{-8}$ |
| Y | max. $\phi$ | $Pd_{37.1}Cu_{55.1}Y_{7.8}$ | $\mathbf{Pd_{48.48}Cu_{43.00}Y_{8.52}}$ | $3.25\times10^{-8}$ |
| | min Pd | $Pd_{35.0}Cu_{54.9}Y_{10.1}$ | $\mathbf{Pd_{45.92}Cu_{43.01}Y_{11.07}}$ | $3.23\times10^{-8}$ |
| | min $\Delta H_{mied}$ | $Pd_{50.0}Cu_{30.0}Y_{20.0}$ | $Pd_{59.09}Cu_{21.17}Y_{19.74}$ | $3.10\times10^{-8}$ |
| La | max. $\phi$ | $Pd_{46.6}Cu_{46.9}La_{6.5}$ | $\mathbf{Pd_{56.09}Cu_{33.70}La_{10.21}}$ | $3.18\times10^{-8}$ |
| | min Pd | $Pd_{35.0}Cu_{45.0}La_{20.0}$ | $Pd_{39.79}Cu_{30.54}La_{29.67}$ | $3.05\times10^{-8}$ |
| | min $\Delta H_{mied}$ | $Pd_{50.0}Cu_{30.0}La_{20.0}$ | $Pd_{53.18}Cu_{19.05}La_{27.77}$ | $3.14\times10^{-8}$ |
| Zn | max. $\phi$ | $Pd_{47.1}Cu_{52.8}Zn_{0.1}$ | $\mathbf{Pd_{59.85}Cu_{40.07}Zn_{0.08}}$ | $1.08\times10^{-8}$ |
| | min Pd | $Pd_{35.0}Cu_{59.6}Zn_{5.4}$ | $\mathbf{Pd_{47.36}Cu_{48.15}Zn_{4.49}}$ | $8.29\times10^{-9}$ |
| | min $\Delta H_{mied}$ | $Pd_{44.4}Cu_{35.6}Zn_{20.0}$ | $Pd_{56.97}Cu_{27.27}Zn_{15.76}$ | $9.13\times10^{-9}$ |

Combining these criteria yields a refined shortlist of candidates for experimental validation, including $Pd_{48.48}Cu_{43.00}Y_{8.52}$, $Pd_{49.08}Cu_{42.45}Sc_{8.47}$, $Pd_{56.09}Cu_{33.70}La_{10.21}$ and $Pd_{52.68}Cu_{40.44}Mg_{6.88}$ (wt%). Overall, Y, Sc, La, and Mg offer the greatest permeability enhancements (up to ~+1.7 − +1.9 × the B2-PdCu archetype), while metallic stabilizers such as Al, Ga, Cr, Fe, Zn and V provide more modest improvements but favorable stability.

In summary, our multi-criteria screening indicates that Pd-lean (compared to B2−$Pd_{60}Cu_{40}$) Pd–Cu alloys stabilized by Y, Sc, La, and Mg constitute the most promising next-generation B2 membranes, balancing cost, stability, and performance.

## 4. Limitations and future outlook

In this study, we trained CatBoost regressors on experimentally observed Pd alloy data. Inevitably, this constrains model generalization, as literature-sourced experimental data are inherently sparse and noisy compared to the highly homogeneous datasets generated by DFT. Additionally, utilizing lattice mismatch as a descriptor required accurate determination of lattice parameters. However, many values were unreported and thus approximated via Vegard's law, introducing systematic errors, particularly for Pd–Cu alloys exhibiting mixed *fcc* and B2 phases. For consistency with prior literature, all virtual alloys were also assumed to possess mixed *fcc*–B2 phases, though their estimated phase ratios may differ *a priori*. As a follow-up, we are currently working on developing a structure-agnostic model.

Furthermore, the Miedema formation enthalpy approximation employed here provides reasonable estimates for ternary solid solutions but is less precise than DFT. Calculating DFT formation energies for the 453,000 virtual alloys in our dataset would require approximating valid structural supercell models for off-stoichiometric compositions and performing numerous single-point calculations, which is computationally prohibitive. Similarly, atomistic ML models are also impractical in this context.

Finally, our ongoing efforts are focused on calculating the enthalpy of formation as well as the adsorption and solution energies for alloy–hydrogen interactions using DFT, followed by fabrication, and measurement of hydrogen fluxes at T ≥ 400 °C to validate our findings.

## 5. Conclusion

We curated a literature dataset of 328 unique Pd-alloy membranes and trained CatBoost regressors to predict hydrogen permeability from composition and operating conditions. To control complexity, we compared raw-feature models with a selection pipeline combining Pearson filtering, fold-wise SHAP-based recursive elimination, and cross-fold union. This reduced MAE/RMSE and yielded compact models. Guided by the one-standard-error rule and parsimony, a 13-feature domain-informed descriptor set preserved accuracy ($R^2 = 0.81$) with far fewer features than alternatives. SHAP explanations are physically consistent: permeability increases with temperature (Arrhenius behavior), lattice expansion relative to Pd, atomic-size mismatch, and alloy stability. We then mapped composition-permeability landscapes for $Pd_{(100-x-y)}Cu_xM_y$ (M ∈ {Mg, Al, Ga, Sc, Cr, Mn, Fe, Ti, Zr, Hf, V, Nb, Ta, Y, La, Zn}) as potential B2 stabilizers. Multi-criteria screening (high predicted permeability, low Pd content, low formation enthalpy) combined with miscibility/segregation filters highlights Pd–Cu–Sc/Y/La and Pd–Cu–Mg as Pd-lean candidates with permeabilities up to ~1.9 times that of B2–$Pd_{60}Cu_{40}$ (wt%). Beyond screening, the model provides a practical tool for estimating permeability for new Pd alloys and approximating activation energies from predicted flux–pressure and permeability–temperature trends. This framework is readily extensible: broader, better-balanced measurements, refined physics-based descriptors, and first-principles stability inputs should improve accuracy and tighten uncertainties. The resulting composition maps and candidate shortlists provide practical reference for designing durable, high-temperature, $H_2$-selective, Pd-lean membranes for separation and membrane reactor applications.

## Data and code availability

The dataset, main code and Supplementary Material to reproduce the results of this study are available at the following GitHub page: https://github.com/AldoKwamibar/Pd-membranes-permeability.

## Acknowledgement

The authors express their gratitude to the Ministry of Education, Culture, Sports, Science, and Technology (Monbukagakusho): MEXT scholarship and Institute of Science Tokyo (formerly Tokyo Institute of Technology) for their financial assistance and support in conducting this research.

## Competing Interest

The authors declare no competing interests.